# LANSCE DIGITAL LOW LEVEL RF UPGRADE*

P. Van Rooy†, M. Prokop, S. Kwon, P. Torrez, L. Castellano, A. Archuleta, C. Marchwinski
AOT-Division, Los Alamos National Laboratory, Los Alamos, New Mexico, USA

## Abstract

Incremental upgrades of the legacy low level RF (LLRF) equipment—50 years for the Los Alamos Neutron Science Center (LANSCE)—involves challenges and problems not seen with new and total replacement opportunities. The digital LLRF upgrade at LANSCE has deployed 30 of the 53 required systems as of September 2022. This paper describes the performance of the digital upgrade, current status, and future installations along with the technical challenges, including unexpected challenges, associated with deploying new digital systems in conjunction with legacy analog equipment. In addition, this paper discusses the operational details of simultaneous multi-energy beam operations using high energy re-bunching, beam-type specific set points and simultaneous multi-beam operations at LANSCE. The adaptability of the digital LLRF systems is essential as the design is able to accommodate new control and beam parameters associated with future systems without significant hardware modifications such as the expected LANSCE Modernization Program. This adaptability of the digital LLRF technology was recently demonstrated with the Module 1, 201.25-MHz high-power RF upgrade completed in 2021.

## LANSCE RF SYSTEM

The LANSCE accelerator, an 800 MeV proton accelerator, saw first light in 1972 and has been in use ever since. The cavity fields were, and in some places still are, controlled by the analog system original to the accelerator. At LANSCE, the LINAC is composed of one 201.25 MHz Drift-Tube LINAC (DTL) sector and seven 805 MHz Coupled-Cavity LINAC (CCL) sectors. 805 MHz sectors are composed of 6~7 CCLs. In contrast, 201.25 MHz sector is composed 4 DTLs and each DTL has its own high voltage system.

In 2014, the LLRF team at LANSCE began installing a LLRF upgrade from the analog systems of the late 1960s and early 1970s to new digital systems [1, 2]. The analog system is becoming more difficult to maintain with increasingly obsolete parts making repairs and maintenance very challenging, a lack of flexibility and the need for an upgraded system as part of an overall LANSCE upgrade. It should be noted that the original analog system was incredibly robust and has been working reliably since 1972.

Currently, 30 of the 53 systems digital LLRF systems have been upgraded and all digital LLRF systems are scheduled to be updated by the end of 2024. As of October 2022, all 201.25 MHz RF systems have been updated and 26 of the 44 805 MHz RF systems have been updated. There are still updates to come for the remaining 805 MHz RF systems, the Low Energy Buncher Transport (LEBT), and the Proton Storage Ring (PSR).

## DIGITAL LOW LEVEL RF UPGRADE

### The System

The new digital LLRF system has more sophisticated control systems, digital processing, and a greatly increased adaptability. The LLRF data is now available over the Experimental Physics and Industrial Control System (EPICS), has a digital I/Q controller, and the flexibility of the system is increased [2, 3].

A Fully Adaptive Feed Forward Iterative Learning Controller has beam implemented for several beam types. The long term robustness of this system and an auto start-up system of the digital LLRF systems are currently under test. Currently in use for the module one 2012.25 MHz systems is a disturbance observer controller.

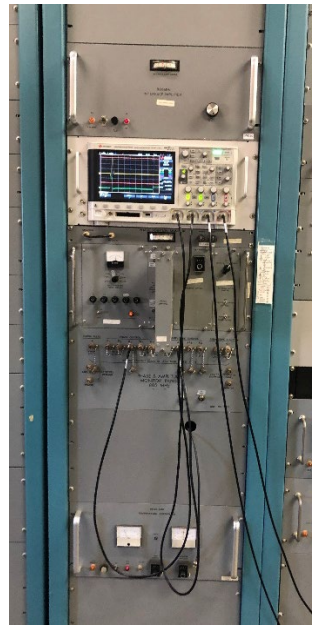

Figure 1: Analog LLRF Rack

### Deployment of digital Low Level RF Systems and integration with existing legacy equipment

The first digital LLRF systems were deployed in conjunction with new high-power (3 MW) 201.25 MHz tube amplifiers in the spring of 2014. Due to integration and control network issues, the deployment was delayed for a year and the first systems were operational in 2015. The final deployment of the remaining 201.25 MHz system happened in spring of 2016. The integration and commissioning with the new tube amplifiers was simplified because both systems were designed with the other in mind.

___________________
*Work supported by the United States Department of Energy, National Nuclear Security Agency, under contract 89233218CNA000001.
†pvanrooy@lanl.gov

Deployment of the digital LLRF to the 1.25-Megawatt klystron-based 805 MHz RF systems was more problematic. This effort required interfacing to and providing 10V gate signals to the legacy systems as well as anticipating future upgrades to the high-power system interface. Since then, a total of 26 805 MHz RF stations are upgraded to the digital LLRF system.

These systems were modified during the annual maintenance outage and commissioned either at LINAC startup or during a 2-day beam development period. Each RF station was set up to operate in either the legacy analog system or the new digital system with the rearrangement of a few cables. This change over was facilitated by the inclusion of independent measurements of RF cavity amplitude and phase so that the current tune was easily replicated with the new system.

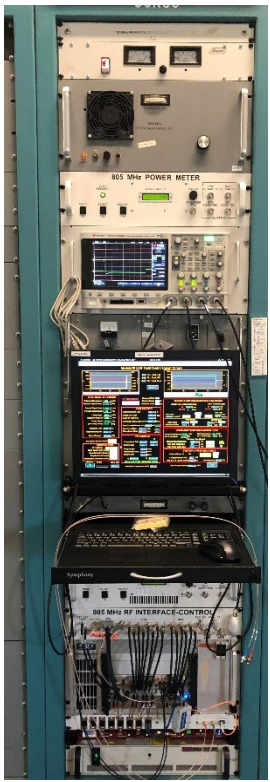

Figure 2: Digital LLRF Rack

### Adaptive Feed Forward Controller

The new feed forward system within the digital LLRF system has individual feed forward settings for each beam type. This offers greater flexibility than the analog LLRF system, which had one feed forward setting for all beam types.

The beam feed forward system uses toroid current detectors to provide a real-time estimate of the beam current and this signal is distributed to each of the LLRF systems. Depending on the location of the module, the systems receive either a sum of H+ and H- current (201.25 MHz RF systems or only H-current (805 MHz RF systems). A static beam feed forward system is adequate for low to moderate beam currents (<15 mA) but high beam currents require additional beam loading compensation for low-spill transport.

To accomplish the tight requirements for low-spill transport, LANSCE developed an adaptive technique to improve the field transient performance of the control system. Using an iterative learning technique to accurately adapt to the particular transient response of the RF cavity for each beam, this controller measures and adjusts the feed forward control signal for minimum field transient. Studies have shown that the majority of beam spill in the LINAC is due to the transient at beam turn-on.

Two types of beam loading compensation feedforward controllers are implemented. The first beam feedforward controller is the static beam feedforward controller (SBFFC), where the detected beam current is read back to the LLRF system and a proper amplification and rotation of the detected beam current generates the feedforward I and Q control signals.

A further improvement is possible by an additional feedforward controller. The second type of the feedforward controller is the iterative learning controller (ILC) [3, 4, 5]. The implemented ILC is the current cycle feedforward (CCF) [4]. That is, the controller output updating rule is

$$u_{k+1} = Q(z)u_k + L(z)e_k + u_{FB} \quad (1)$$

where $u_{FB}$ is the feedback controller and it is given by $u_{FB} = C(z)e_{k+1}$. The reason of this update rule is that the feedforward control is supplementary and the feedback controller is not replaced with ILC and the stability of the closed loop system mainly determined by the PI feedback controller is preserved. It is noted that the steady state error of the system is affected by the Q filter and L filter. In the implementation of ILC, the Q filter is a first order low pass IIR filter and its gain, $q$, is adjustable by the user input through EPICS IOC.

$$Q(z) = q\frac{(1-a)}{z-a}, \quad 0 < a < 1 \quad (2)$$

The pole, $a$, of the Q filter is determined so that Q filter does not deteriorate the closed loop system stability. L filter does not filter out the interesting frequency components of the error signal and should not degrade the closed loop system stability. Because of the trade-off of stability and performance, the L filter, high frequency components of the error if there exist may not be suppressed.

### Frequency Agile Controller

There are a number of control algorithms that can be used to maintain the resonance of an accelerator cavity. The control algorithms often use the transmitted voltage and the reflected voltage in reference to the forward voltage measured with respect to the cavity microwave junction. Each algorithm results in a control output that varies as the resonance frequency of the accelerator cavity varies. The control output indicates the direction and magnitude that the resonance frequency of the cavity must be adjusted to [6, 7].

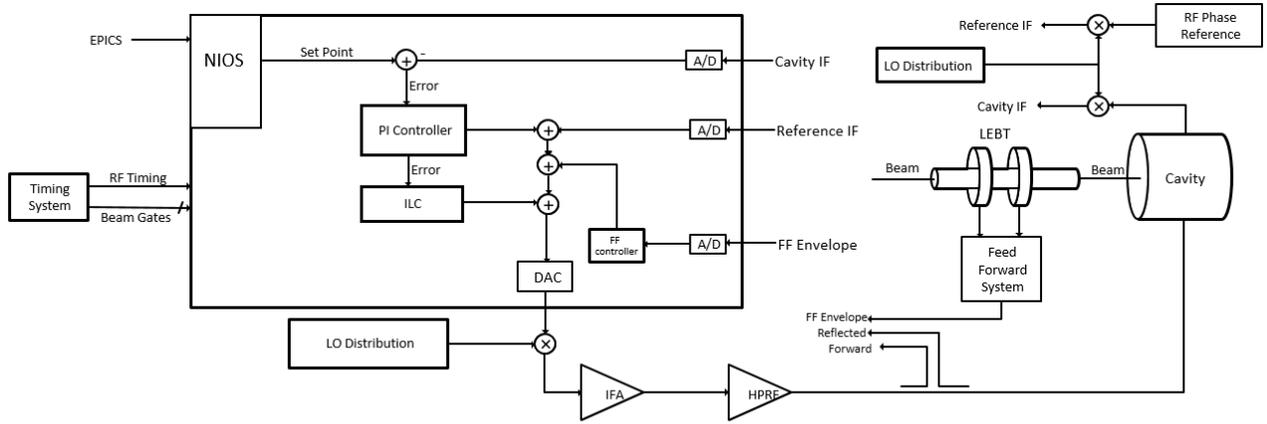

Figure 3: Block Diagram of the digital LLRF system

A preferable resonance control algorithm that gives control linearity, insensitivity to noise, and insensitivity to systematic errors. Easy calculation is based on the imaginary part of the cavity admittance function, which is expressed as I/Q coordinate signals of the cavity filed signal and the forward signal,

$$E = k \frac{I_{cav} \cdot Q_{fwd} - Q_{cav} \cdot I_{fwd}}{I_{cav}^2 + Q_{cav}^2} \quad (3)$$

where k is a gain. E is zero when the resonance frequency of the cavity matches the frequency of the forward signal. E is linear, monotonically increasing function of the detuning frequency. Since E is the error in the cavity resonance, it is applied to a proper controller, such as PI controller, with the proper loop gain to generate the resonance control output.

## Automatic Digital LLRF Turn On

Because of the digitization accompanying the LANSCE Control System (LCS) network, a network based LLRF system control became possible. LLRF system control parameters are assigned with parameter variables (PVs) of the EPICS Database and the operation of the RF system can be performed by adjusting those PVs. MATLAB is used widely in LANCE for analysis of signals and synthesis of LLRF digital signal processing. It is also used for controlling PVs of LLRF system. For this, a MATLAB version of EPICS channel access was developed [8].

The startup procedure of the closed-loop RF operation is comprised to (i) turn on RF; (ii) adjust amplitude and phase set point of the cavity field; (iii) calibrate the open loop gain and phase; (iv) close the PI feedback; (v) calibrate the cavity amplitude and the cavity phase of the external measurements of them to the stored reference values.

The step (iii) is performed to reduce/decouple the crosstalk between I/Q channels and make the overall loop gain be unity. As a result of this calibration, the operating points of the LLRF control system are located close to the amplitude and phase set points and the overdrive of the closed loop system is mitigated.

In LANSCE LLRF system, an additional calibration of step (v) is performed in the closed loop. For the LLRF system, independent, external instruments are deployed to measure the cavity field amplitude and cavity field phase from cavity RF signal with respect to the 201.25MHz or 805 MHz reference RF. The reference values for them are obtained during the beam tuning. These independent amplitude and phase values are valuably used when the operating points of the LLRF, high power RF subsystems, resonance control system, and other systems of LINAC are changed.

Each step of the startup procedure can be performed either manually or automatically. The automatic startup is accompanied with running MATLAB functions the LLRF team coded. The MATLAB functions are run either on MATLAB command window or on Shell Command of the EDM.

## Disturbance Observer Controller

The application of the disturbance observer (DOB) controller is proper to suppress the non-repetitive disturbances at low frequency [9]. The DOB needs the inverse model of the plant (cavity). In general, the plant, the base-band representation of the cavity is a low pass filter characteristics and so the inverse of the plant may be anti-causal, which means the implementation of the inverse of the nominal model, $G_n^{-1}$ on the FCM is impossible. Then, instead of $G_n^{-1}$, $Q_{ob} G_n^{-1}$ is implemented with the design of the disturbance observer $Q$-filter. It is obvious that the filter, $Q_{ob}(s)$ plays a central role in the disturbance observer-based controller. Ideally, to estimate the effect of the disturbance, $Q_{ob}(s)$ should be designed to close to 1 in all the frequency range. However, this may amplify the high frequency sensor/detector noise. Since the plant has lowpass filter characteristics, $Q_{ob}(s)$ is designed as a low pass filter with its relative degree being equal or greater than the relative degree of the plant (model), $G_n$, so that $Q_{ob} G_n^{-1}$ is implementable. The reason that $Q_{ob}(s)$ is a lowpass filter is that the disturbance, d is of low frequency or medium frequency and the sensor/detector noise is usually of high frequency. As a result, the disturbance observer estimates the disturbance of low frequency or medium frequency but rejects sensor/detector noise of high frequency. The cutoff frequency of $Q_{ob}(s)$ is vital in trading off between the stability and the performance, frequency characteristics of the disturbance, d, and the frequency characteristics of senor/detector noise, etc. Higher cutoff frequency yields better disturbance attenuation but it increases the sensitivity to the sensor noise. The DOB outputs are multiplied by adjustable gains and applied to add control signals.

In the LANSCE LINAC, the DOB controller was used to suppress the phase drift of a 20 kW solid-state amplifier that is installed at DTL tank 1 as an intermediate power amplifier [10]. In this case, the nominal plant is modeled with a gain and a two-by-two phase rotation matrix and the phase drift is treated as an additive input disturbance. Since the phase drift is slowly varying, the cutoff frequency of $Q_{ob}(s)$ is set to small. The DOB controller is used further to improve the cavity field error performance for the beam loading. The real beam loading of the LANSCE LINAC is not perfectly repetitive. It has both repetitive and non-repetitive components. The Static beam feedforward controller and iterative learning controller compensate for most of the beam loading effect and the DOB controller improves the error performance further by compensating for the non-repetitive beam loading components.

## ADAPTABILITY OF THE DIGITAL LOW LEVEL RF UPGRADE

The digital Low Level RF systems at LASNCE were designed with adaptability in mind. The adaptability of the FPGA-based digital LLRF system came to the forefront with the new 201.25 MHz tube amplifiers. During the commissioning of the systems, it was discovered that too much reflected power was impinging on the output of the tube. Since there were not any circulators to absorb the reflected power, the solution was to provide a ramped RF drive for the high-power system. With the FPGA-based design we were able to optimize the ramp-up time for each of the specific cavities and to minimize the settling time of the cavity field. This programmability also allowed us to provide the specific signals to enable additional system protection capability.

The commissioning of a new 400-kilowatt RF system for Module 1 encountered an unusual behavior with the solid-state water-cooled driver amplifier for the final output tube amplifier. Because of the temperature change in the amplifier from the steady state operation, the phase through the system changed over 30° and impacted the performance of the RF control system. To compensate for this, we added a disturbance observer that counteracted the phase change independent of the overall control system. This solution was easily implemented in the FPGA and used a spare RF input for the disturbance source. This was implemented and tested in the 3-week commissioning phase for the amplifier system.

The frequency dependent components are located in one chassis. LANSCE has several frequencies within the LLRF systems, including 2.8 MHz, 16.77 MHz, 25.15625 MHz, 176.09375 MHz, 201.25 MHz, 779.84375 MHz, and 805 MHz. By having one chassis contain the frequency dependent circuits, it makes adapting to a different frequency straightforward as there is only one chassis and a few components in the tunnel that need to change.

## SIMULTANEOUS MULTI-ENERGY BEAM OPERATIONS

LANSCE hosts 5 main user facilities (IPF, PRAD, UCN, Lujan and WNR) that require differing rates of beam delivery multiplexing the 120 Hz pulse rate with their specific beam. This is handled by the distributed timing system control of various pulse power equipment dedicated to the redirection of the beam to the various areas. In addition to temporal distribution, LANSCE has the capability to operate at different energies on a pulse-to-pulse basis (120 Hz) by delaying individual RF stations. LANSCE has delivered beams from 256 MeV up to 800 MeV with gradation of approximately 10 MeV. To enhance this capability, the RF modules are able to also perform a bunching capability for the low energy beams to maintain beam integrity throughout the long drift spaces of the LINAC. For bunching each RF control system will have a beam unique amplitude and phase settings for bunching versus the normal accelerator operation. This bunching functionality is implemented in the FCM and the temporal control is in the HPM.

## CONCLUSION

LANSCE is just over half way complete with the digital LLRF RF upgrade. This upgrade has greatly improved the flexibility and adaptability of the system. The remaining systems to install include the Proton Storage Ring, the Low Energy Beam Transport and the remaining 805 MHz systems.